\documentclass[final,3p,times,twocolumn]{elsarticle}

\usepackage{graphicx}    
\usepackage{dcolumn}     
\usepackage{bm}          
\usepackage{color}       
\usepackage{multirow}
\usepackage{enumitem}
\usepackage{units}
\usepackage{url}
\usepackage{amsmath}
\usepackage{mathrsfs}

\newcounter{bla}

\journal{Computer Physics Communications}

\begin{document}

\begin{frontmatter}



\title{Histogram-Free Multicanonical Monte Carlo Sampling to Calculate the Density of States}

\author[alfred]{Alfred C.K. Farris \corref{cor1}}
\ead{alfred.farris@uga.edu}
\address[alfred]{Center for Simulational Physics, Department of Physics and Astronomy, The University of Georgia, Athens, GA 30602, U.S.A.}
\cortext[cor1]{Corresponding authors.}

\author[ornl]{Ying Wai Li \corref{cor1}}
\ead{yingwaili@ornl.gov}
\address[ornl]{National Center for Computational Sciences, Oak Ridge National Laboratory, Oak Ridge, Tennessee 37831, U.S.A. \fnref{doe_declaimer}}
\fntext[doe_declaimer]{This manuscript has been authored by UT-Battelle, LLC 
under Contract No. DE-AC05-00OR22725 with the U.S. Department of Energy. The 
United States Government retains and the publisher, by accepting the article 
for publication, acknowledges that the United States Government retains a non-
exclusive, paid-up, irrevocable, worldwide license to publish or reproduce the 
published form of this manuscript, or allow others to do so, for United States 
Government purposes. The Department of Energy will provide public access to 
these results of federally sponsored research in accordance with the DOE Public 
Access Plan (http://energy.gov/downloads/doe-public-access-plan).} 

\author[ornl]{Markus Eisenbach}

\begin{abstract}
We report a new multicanonical Monte Carlo algorithm to obtain the 
density of states for physical systems with continuous state variables 
in statistical mechanics.  Our algorithm is able to obtain a closed-form 
expression for the density of states expressed in a chosen basis set, 
instead of a numerical array of finite resolution as in previous variants of
this class of MC methods such as the multicanonical sampling and Wang-Landau 
sampling. This is enabled by storing the visited states directly and avoiding 
the explicit collection of a histogram. This practice also has the advantage 
of avoiding undesirable artificial errors caused by the discretization and 
binning of continuous state variables. Our results show that this scheme is 
capable of obtaining converged results with a much reduced number of Monte 
Carlo steps, leading to a significant speedup over existing algorithms.
\end{abstract}


\end{frontmatter}

{\bf PROGRAM SUMMARY}

\begin{small}

\noindent
{\em Program Title: HistogramFreeMUCA}                        \\
{\em Licensing provisions: BSD 3-clause}                      \\
{\em Programming language: Python}                            \\
{\em Nature of problem:}                                      \\
This program implements a novel algorithm to obtain the density of states of a 
physical system, expanded in a chosen basis set. Unlike existing algorithms 
that return the density of states as a numerical array, this algorithm avoids 
binning of a continuous variable and is able to express the density of states 
as a closed-form expression. It is thus suitable for the study of the 
statistical mechanics and thermodynamic properties of physical systems where 
the density of a continuous state variable is of interest. \\
\noindent
{\em Solution method:}\\
The new algorithm presented here is a special reweighting method in classical 
Monte Carlo approaches. In particular, it can be regarded as a descendant and a 
hybrid method closely related to the multicanonical method and Wang-Landau 
sampling. \\
{\em Additional comments:}\\
Most updated source code can be found at: 
\url{https://github.com/yingwaili/HistogramFreeMUCA}

\end{small}

\section{Introduction}

Monte Carlo (MC) methods are one of the major computational techniques
in statistical physics for the study of finite temperature properties and
thermodynamics of materials \cite{landau_binder_2014}. Traditional MC
methods such as the Metropolis algorithm \cite{metropolis_equation_1953}, an
importance sampling method, works by generating a Markov chain of
energy states $E$ that obey the Boltzmann distribution, $e^{-E/k_B T}$, 
which describes the probability of finding the system at a certain
energy state at a given temperature $T$. Thermodynamics properties are
then calculated by averaging over the entire Markov chain after
equilibration. A well-known limitation of the Metropolis method is the
``critical slowing down'' near phase transitions \cite{hohenberg_theory_1977},
where the correlation time diverges at the critical temperature
$T_C$. Hence, simulations around and below $T_C$ are simply impractical or
unreliable to perform.

Important breakthroughs were introduced by advanced techniques
such as the reweighting methods, which allow for the procurement of a
distribution function of properties. They can be used to obtain
properties at a temperature other than the simulation temperature by
``reweighting'' the distribution function properly: umbrella sampling
\cite{bennett_efficient_1976, torrie_nonphysical_1977}, multihistogram method
\cite{ferrenberg_optimized_1989}, multicanonical (MUCA) sampling
\cite{berg_multicanonical_1991, berg_multicanonical_1992}, and more recently
Wang-Landau (WL) sampling \cite{wang_efficient_2001, wang_determining_2001}, 
all belong to this class of reweighting methods. Because of a special 
formulation of the sampling weights that control the acceptance probability, 
the random walks in these methods are not ``trapped'' in local minima as
in Metropolis sampling. They are thus able to circumvent the critical
slowing down problem. Among the reweighting methods, Wang-Landau
sampling is proven to be quite robust because the simulation is
performed \textit{independent} of temperature. The resulting distribution
function is essentially the density of states (DOS), i.e.,
the energy degeneracy of the system. Thus it reflects only the intrinsic
properties defined by the Hamiltonian. The DOS allows for direct
access to the microcanonical entropy, with which all the
thermodynamics properties including the specific heat and free energy
can be calculated. This feature is essential to enable a reliable
study of phase transitions and critical phenomena, particularly at low
temperatures. 

With the advancement of high performance computers (HPC), it is now possible
to combine Wang-Landau sampling with first-principles methods,
e.g. density functional theory (DFT) \cite{hohenberg_inhomogeneous_1964, 
kohn_self-consistent_1965}, to simulate finite temperature materials properties 
to a high accuracy that is comparable with experimental observations
\cite{eisenbach_scalable_2009, eisenbach_first_2011, khan_density-functional_2016}. 
However, first-principles energy calculations are 
computationally intensive; and yet a reliable Wang-Landau sampling often needs 
a minimum of millions of MC steps (i.e. energy calculations) for one single 
simulation. The time required to finish a simulation is often measured in weeks 
or even months on one of the fastest supercomputers currently available. Such a 
huge computational cost is barely affordable. The type of scientific
problems that can be practically solved by this approach are, for this
reason, still very limited. 

To address this problem, improvements of existing Monte Carlo
algorithms are required. In general, two feasible strategies are available: one
is the parallelization of existing algorithms, in which computational
cost is spread over multiple computing units. Examples include
parallel tempering \cite{swendsen_replica_1986, hukushima_exchange_1996}, 
parallel Wang-Landau sampling on a graphical processing unit (GPU)
\cite{yin_massively_2012}, replica-exchange Wang-Landau sampling
\cite{vogel_generic_2013,vogel_scalable_2014}, and parallel multicanonical 
sampling \cite{zierenberg_scaling_2013, gross_massively_2018}. Another strategy 
is to find ways to reduce the number of MC steps needed to complete a 
simulation. This is normally done by introducing tricks within the framework of 
existing algorithms; but the number of MC steps saved is often small.

In this work, we present a new multicanonical Monte Carlo algorithm
that takes both strategies into account. Our scheme is readily
parallelizable as in \cite{zierenberg_scaling_2013,gross_massively_2018} to 
exploit the power of current HPC architectures. In addition, our algorithm is 
able to attain comparable accuracy with Wang-Landau sampling, using only 
about 1/10 of the number of MC steps. This order of magnitude reduction 
in the number of energy evaluations is particularly crucial when 
first-principles methods are employed for calculating the energy. 
Moreover, for the very first time, our algorithm provides a viable means 
to obtain the density of states as a closed-form expression.  The utility of 
this is twofold: Firstly, working with a closed-form expression during the 
simulation allows us to avoid using histograms to represent continuous 
variables, and in turn bypass systematic errors associated with the choice of 
the resolution for the order parameters, i.e., the bin width of the histogram 
or density of states as in multicanonical and Wang-Landau sampling. Thus, this 
framework will be of particular use when studying statistical mechanical models 
with continuous phase-space variables, e.g., the classical Heisenberg model, 
coarse-grained biopolymer models, Lennard-Jones fluids or clusters, which will 
be discussed in follow-on publications. Secondly, because we are able to obtain 
a closed-form expression, this algorithm will be particularly useful
to fit the functional form of the density of states to aid theoretical studies.

This paper is extended based on the proceeding paper accepted in the 
Platform for Advanced Scientific Computing Conference (PASC'17) 
\cite{li_histogram-free_2017}. Here, we include a Python code to illustrate the 
implementation of our algorithm. We also added Appendix A to clarify the 
definitions and the use of the Kolmogorov-Smirnov test employed in our scheme, 
as it is a major factor affecting the behavior and the accuracy of the
algorithm presented in this work.

\section{Description of the algorithm}
\label{algorithm}

\subsection{An overview}
Our novel algorithm is inspired by previous multicanonical (MUCA)
\cite{berg_multicanonical_1991,berg_multicanonical_1992} and Wang-Landau
(WL) \cite{wang_efficient_2001,wang_determining_2001} Monte Carlo methods. 
Therefore our algorithm shares many of its underlying principles with these 
earlier methods. The major advantage of our scheme over the previous ones is
that our algorithm, for the first time, provides a viable avenue to estimate a 
closed-form expression of the density of states in energy, denoted by $g(E)$. 
Here $E$ stands for an energy the simulated physical system can realize. We 
assume an expansion for the natural log of $g(E)$ in terms of an orthonormal 
basis set $\{\phi_i(E)\}$ with each term weighted by the coefficient $g_i$:
\begin{equation}
\label{g(E)}
\ln g(E) = \sum_{i=1}^{N}g_i\phi_i(E),
\end{equation}
with $N$ being the number of basis functions utilized in the
expansion. The estimation of $g(E)$ will be improved iteratively
later during the course of the simulation by a similarly defined, yet
slightly modified, correction function $c(E)$:
\begin{equation}
\label{correction1}
\ln c(E) = \sum_{i=1}^{N}c_i\phi_i(E),
\end{equation}
where $c_i$ is the weighting coefficient for $\phi_i(E)$ in the
correction.

The algorithm begins with an initial guess of $\tilde{g}(E) = 1$
(i.e., $\ln \tilde{g}(E) = 0$). In other words, it is a uniform
distribution. Next, a series of Monte Carlo moves is performed 
and a Markov chain of $k$ energies is generated to construct a 
data set $\mathcal{D}=\{ E_1, E_2, ..., E_j, ..., E_k \} $ according 
to the following acceptance probability:
\begin{equation}
\label{acc_prob}
p(E_j \to E_{j+1}) = \min\left( \frac{\tilde{g}(E_j)}{\tilde{g}(E_{j+1})} , 1 \right ).
\end{equation}
Note that the acceptance rule follows that of the Wang-Landau
algorithm \cite{wang_efficient_2001}. That is, if the trial energy $E_{j+1}$ is
rejected, the previous accepted state of the system should be recovered,
but the associated energy $E_j$ would be counted again as $E_{j+1}$. A
Monte Carlo move is then performed on the reverted state to generate
the next trial energy $E_{j+2}$.

After the data set $\mathcal{D}$ is generated, it is used to find the
correction $c(E)$ that improves the estimated density of states
$\tilde{g}(E)$ such that:
\begin{equation}
\label{updateDOS}
\ln \tilde{g}(E) \to \ln \tilde{g}(E) + \ln c(E).
\end{equation}
The details of obtaining the correction function $c(E)$ from the data set
$\mathcal{D}$ will be further described below in subsection
\ref{correction}. For now, assume that we have updated the estimated
density of states $\tilde{g}(E)$ with $c(E)$ using
Eq. (\ref{updateDOS}). The simulation is then brought to the next
iteration with $\mathcal{D}$ and $\ln c(E)$ reset to empty or zero,
respectively, while $\tilde{g}(E)$ will be kept unchanged and carried
over to the next iteration as the new sampling weights. The process of
generating the data set $\mathcal{D}$ and obtaining the correction
$c(E)$ is then repeated. The iteration repeats and terminates when
$\ln c(E) \rightarrow 0$. The DOS of the system is a fixed point
of the iterative process when convergence is reached.

\subsection{Obtaining the correction $c(E)$ from data set
  $\mathcal{D}$}
\label{correction}

The key of the above framework is to obtain a closed-form expression for the 
correction $c(E)$, or $\ln c(E)$ in the actual implementation of our algorithm. 
To do so, we must first obtain such an expression for the empirical cumulative 
distribution function (ECDF) of the data $\mathcal{D}$, from which $c(E)$ can 
be deduced. 

\subsubsection{Obtaining a closed-form expression for the empirical cumulative distribution function (ECDF)}

We construct the ECDF following the scheme proposed by Berg
and Harris \cite{berg_data_2008}, which we outline here. Recall that
our data set $\mathcal{D}$ is a collection of $k$ energies generated
from a Monte Carlo Markov chain. The energies are first sorted in
ascending order: 
\begin{equation}
\begin{aligned}
\mathcal{D} &= \{ E_1, E_2, ..., E_j, ..., E_k \} \\
                    &= \{ E_{\pi_1}, E_{\pi_2}, ..., E_{\pi_j}, ..., E_{\pi_k} \},
\end{aligned}
\end{equation}
where $\pi_1$, ..., $\pi_k$ is a permutation of 1, ..., $k$ such that
$E_{\pi_1} \leq E_{\pi_2} \leq ... \leq E_{\pi_j} \leq ... \leq E_{\pi_k}$. If all $E_{\pi_j}$ are different, the empirical cumulative distribution function (ECDF) is defined as:
\begin{equation}
\bar{F}(E) = \frac{j}{k},  \hspace{5mm}\textrm{for $E_{\pi_j} \le E < E_{\pi_{j+1}}$}.
\end{equation}

\noindent
If multiple $E_{\pi_j}$'s have the same value, e.g., $E_{\pi_j} = E_{\pi_{j+1}} = ... = E_{\pi_{j+l}}$, the ECDF would take a larger step:
\begin{equation}
\bar{F}(E) = \frac{j+l}{k}, \hspace{1mm} \textrm{for $E_{\pi_j} = E_{\pi_{j+1}} = ... = E_{\pi_{j+l}} \le E < E_{\pi_{j+l+1}}$}.
\end{equation}
Assuming that the ECDF can be decomposed into two components:
\begin{equation}
\label{ECDF}
\bar{F}(E) = F_0(E) + \bar{R}(E),
\end{equation}
where $F_0(E) = (E-E_{\pi_1}) / (E_{\pi_k}-E_{\pi_1})$ is a straight 
line for $E \in [E_{\pi_1}, E_{\pi_k} ]$, and $\bar{R}(E)$ defines the
empirical remainder. The choice of $F_0(E)$ as a straight line is
based upon the following observations: for traditional histogram
methods, the ECDF plays the role of the cumulative histogram that can
be constructed directly from the histogram $H(E)$. Nevertheless, in 
the continuous limit, the ECDF does not suffer from the
binning effect. The derivative of the ECDF is then equivalent to the
histogram in traditional methods: $H(E) = d\bar{F}(E) / dE$. In such
schemes, obtaining a ``flat'' histogram is an indicator that the
energy space is being sampled uniformly. The sampling weights are
continuously adjusted to direct the random walk from highly accessible
states to rare events, either periodically in MUCA or adaptively in
WL, to achieve this goal. Here, a ``flat histogram'' is equivalent to
an ECDF with a constant slope.

The next task is to find a closed-form expression for the remainder $R(E)$ to 
fit the empirical data $\bar{R}(E)$. $R(E)$ signifies the deviation from the 
ideal (uniform) sampling, which will inform us on how to amend the weights to 
drive the random walks. It is expected that $R(E)$ will be related to the 
correction $c(E)$. Therefore, it is reasonable to assume that $R(E)$ can be
similarly expanded in terms of an orthonormal basis set $\{\psi_i(E)\}$:
\begin{equation}
\label{R(E)}
R(E) = \sum_{i=1}^{m}r_i\psi_i(E),
\end{equation}
where $m$ is the number of terms in the expression. The coefficients
$r_i$ can be then be found by:
\begin{equation}
\label{remainder_coef}
\begin{aligned}
r_i &= \mathcal{N} \int_{E_{\pi_1}}^{E_{\pi_k}} R(E) \psi_i(E) dE,
\end{aligned}
\end{equation}
with $\mathcal{N}$ being a normalization constant dependent on the
choice of the basis set $\{\psi_i(E)\}$. Note also that the basis set
$\{\psi_i(E)\}$ needs to be able to satisfy the ``boundary conditions''
at $E_{\pi_1}$ and $E_{\pi_k}$ that $R(E_{\pi_1}) = R(E_{\pi_k}) = 0$,
by definition. Since $R(E)$ is indeed an empirical function resulted
from the ECDF, the integral in Eq. (\ref{remainder_coef}) is a 
simple numerical summation.

The remaining question is to determine the number of terms $m$ in
Eq. (\ref{R(E)}) to fit $\bar{R}(E)$ while avoiding overfitting. This 
is done by an iterative procedure starting from $m = 1$ where there 
is only one term in the sum. A statistical test is then performed to 
measure the probability $p$ that this $R(E)$ is a ``good'' fit to 
$\bar{R}(E)$. That is, $p$ is the probability of obtaining the 
empirical remainder $\bar{R}(E)$ if the data is generated according 
to the distribution specified by $R(E)$. We follow the suggestion of 
\cite{berg_data_2008} and use the Kolmogorov-Smirnov test 
\cite{KolmogorovAN1933, Smirnoff1939} (see Appendix A for details), but other 
statistical tests for arbitrary probability distributions can also be 
used. If $p < 0.5$, we increase $m$ to $m+1$ and repeat the statistical 
test, until $p \geq 0.5$ is reached. The number of terms $m$ is then 
fixed at this point.  Note that in principle, increasing $m$ further 
would result in a ``better fit'' and thus a larger $p$. However, it is 
not preferable because it increases the risk of over-fitting a 
particular data set and would be difficult to correct through latter 
iterations. Thus we choose the criterion $p \geq 0.5$ to keep the 
expression as simple as possible, and to maintain some levels of 
stability against noise.

With the expression of $R(E)$, a closed-form approximation of the ECDF can then be obtained:
\begin{equation}
\label{F(E)}
F(E) = F_0(E) + R(E).
\end{equation} 

\subsubsection{From ECDF $F(E)$ to the correction $c(E)$}
                                                    
Finally, the expression of $F(E)$ in Eq. (\ref{F(E)}) is used to
obtain the correction $c(E)$ (or $\ln c(E)$ in practice). Recall the
definition of the cumulative distribution function (CDF) for a
continuous variable, which can be constructed from the probability
density function. They are, respectively, equivalent to $F(E)$ and
$H(E)$:
\begin{equation}
\label{CDF}
F(E) = \int_{-\infty}^{E} H(E') dE'.
\end{equation} 

Combining Eqs. (\ref{F(E)}) and (\ref{CDF}) and taking derivatives
of both sides to obtain $H(E)$ yields:
\begin{equation}
\begin{aligned}
\label{H(E)}
H(E) = \frac{dF(E)}{dE} &= \frac{dF_0(E)}{dE} + \frac{dR(E)}{dE} \\
                                   &= \frac{1}{E_{\pi_k}-E_{\pi_1}} + \sum_{i=1}^{m}r_i \frac{d\psi_i(E)}{dE}.
\end{aligned}
\end{equation}

As in traditional multicanonical sampling methods, the histogram
$H(E)$ is used to update the estimated density of states $\tilde{g}(E)$,
hence the sampling weights for the next iteration. Observe that the
first term in Eq. (\ref{H(E)}) is just a constant independent of the
value of $E$, it can be safely omitted in the correction. Thus, we choose 
to write the log of the correction using part of the histogram (taking 
only the second term in Eq. (\ref{H(E)})):
\begin{equation}
\begin{aligned}
\label{correction2}
\ln c(E) &= \sum_{i=1}^{m}r_i \frac{d\psi_i(E)}{dE},
\end{aligned}
\end{equation}
which, in our special case, has the same form as Eq. (\ref{correction1}) with 
\begin{equation*}
c_i \phi_i(E) = r_i \frac{d\psi_i(E)}{dE} \hspace{3mm} \textrm{  and  } \hspace{3mm} N = m.
\end{equation*}
In general, we need to ensure that $\frac{d\psi_i(E)}{dE}$ can be
expanded in terms of $\{\phi_j(E)\}$, i.e., 
\begin{equation*}
\frac{d\psi_i(E)}{dE} = \sum_j p_{ij}\phi_j(E). 
\end{equation*}
Finally, the estimated density of states $\tilde{g}(E)$ is updated
using Eq. (\ref{updateDOS}). 

\subsection{A note on the update of the density of states}
There is a major difference between our scheme and the traditional MUCA algorithm in the way the density of states $\tilde{g}(E)$ gets updated. Effectively, our scheme updates $\tilde{g}(E)$ using $\exp(\Delta H(E))$ (where $\Delta H(E)$ denotes the deviation of the histogram $H(E)$ from uniform sampling, i.e., a ``flat'' histogram), whereas traditional MUCA updates $\tilde{g}(E)$ using the histogram $H(E)$ itself directly. Our scheme is therefore more responsive and proactive in adjusting the sampling weights, and thus $\tilde{g}(E)$, to guide the random walker to achieve uniform sampling. 

\section{Test case: numerical integration}
\label{integration}

The algorithm was originally designed with the motivation of sampling
physical systems with a continuous energy domain. Yet, as the majority
of these systems do not have a closed-form solution, it is difficult
to quantify the accuracy of the algorithm. We thus apply it to perform
numerical integration using the scheme suggested by
Ref. \cite{li_numerical_2007} as a proof-of-principle.

Note, however, that our method is not meant to be an efficient
algorithm for performing numerical integration. As pointed out in
Ref. \cite{li_numerical_2007}, there is a one-to-one correspondence
between numerical integration and simulating an Ising model when put
under the Wang-Landau sampling framework. This applies to our
algorithm too and as long as we choose an integrand that is continuous
within the interval $[y_{\min}, y_{\max}]$, it is equivalent to the
situation of having a continuous energy domain for a physical
system. Moreover, numerical integration is indeed a more stringent
test case for our algorithm (and other histogram MC methods such as
Wang-Landau sampling in general), because the ``density of states''
$g(y)$ is usually more rugged than the density of states of a real
physical system.

If one can find an expression for the normalized $g(y)$, which
measures the portion of the domain within interval $[a, b]$
corresponding to a certain value of $y$, then the integral
can be found by summing the ``rows'' up (multiplied by the value of
$y$) instead of the columns in the following manner:
\begin{equation}
I = \int_a^b y(x) dx = \int_{y_{\min}}^{y_{\max}} g(y) y dy .
\end{equation}

Note that $g(y)$ needs to be normalized such that
\begin{equation}
\int_{y_{\min}}^{y_{\max}} g(y) dy = b-a.
\end{equation}

\begin{figure}[ht!]
\centering
\includegraphics[height=2in, width=0.85\columnwidth]{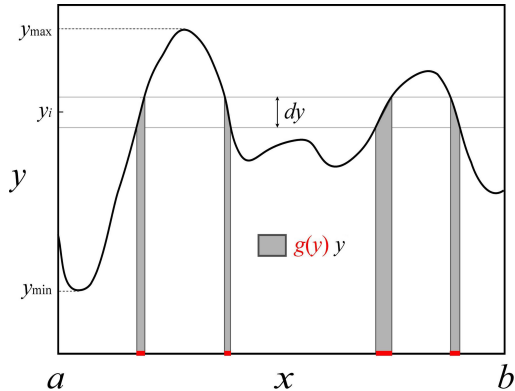}
\caption{A schematic diagram showing the numerical integration
  notations. The red regions on the $x$-axis marked the portion within
  the interval $[a, b]$ that gives a certain value $y$. All the areas
  shaded in grey add up to give $g(y)y$. }
\end{figure}

We apply our algorithm to perform the following integration where the
exact integral is known:
\begin{equation}
I = \int_{-2}^2 x^2 dx = \frac{16}{3} = 5.33333 \cdots ,
\end{equation}
and the ``density of states'' $g(y)$ can be expressed analytically:
\begin{equation}
g(y) = \frac{2(2-\sqrt{y})}{y}  \hspace{5mm}\textrm{for $y > 0$}.
\end{equation}

We use a Fourier sine series as the basis set $\{\psi(E)\}$ to fit the
remainder $R(E)$, and therefore a Fourier cosine series as the basis set
$\{\phi(E)\}$ for constructing the correction $\ln c(E)$ and updating
the density of states $\ln g(E)$. Moreover, we employ the Kolmogorov-Smirnov
test in the $R(E)$ fitting step to determine if the expression obtained is a
good fit to the dataset $\mathcal{D}$, using a criterion of $p =
0.5$. The experiment is done for different numbers of data in
the data set, with $k =$ 250, 500, 1000 and 2000.

\begin{figure}[h!]
\centering
\includegraphics[width=0.95\columnwidth]{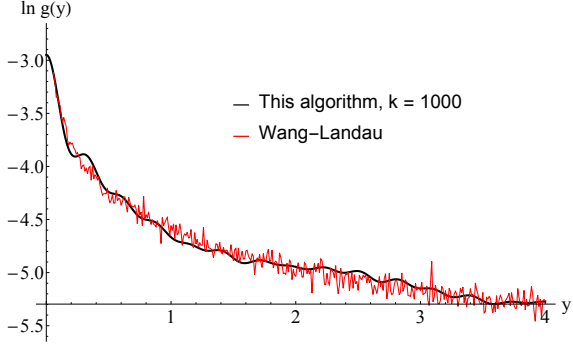}
\caption{Density of states $g(y)$ at the $120^\textrm{th}$ iteration,
  obtained using 1000 data points in a data set (black curve), a total
  of $1.2 \times 10^5$ MC steps are used. It is compared to a final
  $g(y)$ obtained using Wang-Landau sampling (red curve); this
  particular run requires $1.1 \times 10^6$ MC steps to complete. The
  DOS obtained from our algorithm is significantly smoother, yet its
  fluctuations fall within the statistical noise of the Wang-Landau
  DOS.}
\label{normDOS}
\end{figure}

We note that the Fourier sine and cosine series are not well suited
basis sets for this problem due to their oscillatory properties. Yet
the algorithm works surprisingly well. In Figure \ref{normDOS},  we
show a resulting density of states, $g(y)$, compared to that obtained
using Wang-Landau sampling. The fluctuations of our $g(y)$ fall within
the statistical noise of the WL density of states. 

\begin{figure}[ht!]
\centering
\includegraphics[width=0.95\columnwidth]{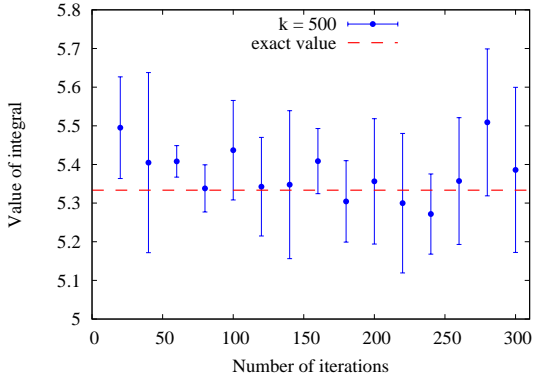}
\caption{Integral of $x^2$ over $x \in [-2,2]$, obtained using 500
  data points in a data set. Error bars are obtained from five
  independent runs.}
\label{k500}
\end{figure}

\begin{figure}[ht!]
\centering
\includegraphics[width=0.95\columnwidth]{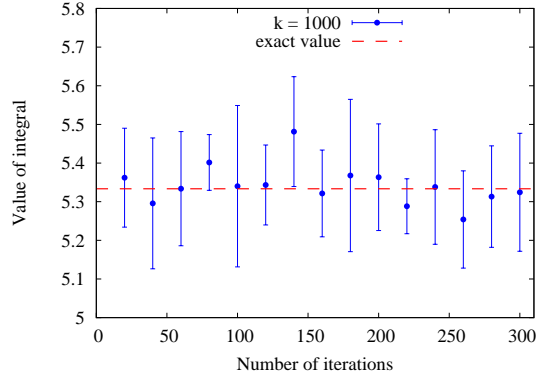}
\caption{Integral of $x^2$ over $x \in [-2,2]$, obtained using 1000
  data points in a data set. Error bars are obtained from five
  independent runs.}
\label{k1000}
\end{figure}

The values of the estimated integral at different iterations for $k =
500$ and $k = 1000$ are shown in Figures \ref{k500} and \ref{k1000},
respectively.

\begin{figure}[ht!]
\centering
\includegraphics[width=0.95\columnwidth]{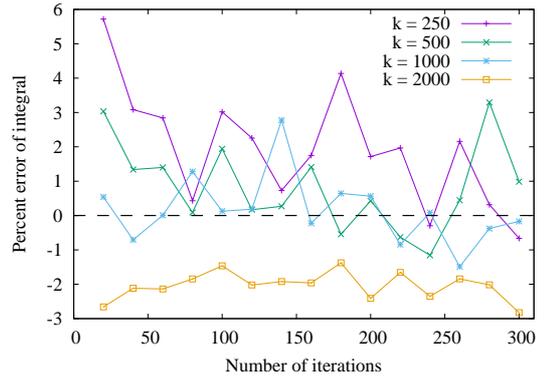}
\caption{Percent errors of the averaged estimated integrals at
  different iterations using various numbers of data points $k$ in a
  data set. The integrals for all cases are averaged over five
  independent runs. Except for $k=2000$, all other cases converge to
  the exact value eventually with small errors (within $\pm1\%$).}
\label{errors}
\end{figure}

We observe that the number of data points $k$ in a data set within an
iteration plays an important role in the accuracy. Both under-fitting
from insufficient data and over-fitting from excessive data would
produce inaccurate results. In both Figures \ref{k500} and
\ref{k1000}, most estimated integrals agree with the exact value to
within the error bars. No systematic correlation with the number of
iterations is observed for either the estimated values of the integral or 
the magnitude of the error bars. Using $k=500$ or $k=1000$ does not seem to result in
significant  differences in the estimated value of the
integral. However, if we extend the studies and use fewer or more data
points in the data set $\mathcal{D}$, we observe different behavior as
shown in Figure \ref{errors}. For the $k = 250$ case, the integral is slightly 
overestimated at the first 200 iterations or so. The
percent errors fall back to within the same ranges as in the $k=500$ and
$k=1000$ cases later. This is reasonable because as the number of
iterations increases, more data are taken to correct the estimated
density of states.

However, the integral is, unexpectedly, systematically
underestimated for the $k = 2000$ case. We also observe that the
number of terms in the expression of $R(E)$ and eventually $\ln g(E)$
generally increases with the number $k$ (Table \ref{nterms}).  A
larger number of data results in a more detailed fitting of the
DOS, hence more terms are used in the construction of the correction.
Unfortunately, there is also a higher risk of fitting the noise
``too well'', causing an over-fitting of the data set (this is due to 
the dependence of the KS test on the number of data points in a data set - 
see Appendix A for more details). On the other hand, using too few data 
points (such as $k = 250$) results in larger fluctuations in the values 
of the integral as well as in the number of terms $N$ in the expression. 
From our observations, using about $k=1000$ data points in a data set is 
the safest and it strikes a good balance between under-fitting (or even 
mal-fitting) and overfitting.

\begin{table}
\centering
\begin{tabular}{|c|c|} \hline
$k$   &  $N$ \\ \hline
250   &  $19.4 \pm 14.6$ \\ \hline
500   &  $17.8 \pm 6.1$   \\ \hline
1000 &  $29.6 \pm 4.3$   \\ \hline
2000 &  $58.2 \pm 13.0$ \\
\hline\end{tabular}
\caption{\label{nterms} Averaged number of basis functions ($N$) for
  different number of data points ($k$) in each data set. The average
  values and errors are obtained from five independent runs.  The number
  $N$ is determined within the algorithm, at which the statistical
  test reaches $p \geq 0.5$. }
\end{table}

Note that the above experiments complete within hundreds of
iterations. Considering $k=1000$ data points in an iteration, the
total number of MC steps needed is of the order of
$10^5$. Comparing to the order of $10^6$ MC steps in Wang-Landau
sampling, our scheme is more efficient and it saves about 10$\times$
MC steps.  The reason is that when we correct the estimated DOS (i.e.,
sampling weights),  the correction is constructed to drive the random
walk \textit{intentionally} to achieve uniform sampling, or a ``flat
histogram'', as opposed to an incremental correction using the
histogram as in MUCA or WL sampling. We believe that our correction
scheme can also be applied to simple models with discrete energy
levels and yields significant speedup. 

\section{An improved scheme for better convergence}
\label{improved_algorithm}

While the results above showed that our proposed scheme is successful,
one problem is that it is still difficult to determine whether
convergence has been reached. Here, we suggest a possible way to
improve the quality of the results with two slight modifications to
the original scheme.

Firstly, when determining the number of terms for the
remainder $R(E)$ ($m$ in Eq. (\ref{R(E)})), the original scheme starts
from $m=1$ and increments it to $m+1$ sequentially until the
statistical test gives a score of $p \geq 0.5$. We observe that this
practice very often results in the update of the first few
coefficients only.  A remedy to it is that after the number $m$ is
determined in the first iteration, in the later iterations we propose
random permutations of the terms for the expansion rather than 
adding the terms sequentially. The statistical test 
is still used to terminate the iteration as before. This way, every 
coefficient will have a roughly equal chance to get updated and refined.

Secondly, since the correction in Eq. (\ref{correction2}) will drive the
random walker in a way to achieve uniform sampling, we observe that it
is beneficial to use a milder correction update to drive the random
walker at a smaller step at a time. To do so, we rewrite Eq. (\ref{correction2})
with a pre-factor $s$ to take only a portion of $R(E)$ as
the correction:
\begin{equation}
\begin{aligned}
\label{correction3}
\ln c(E) &= s \sum_{i=1}^{m}r_i \frac{d\psi_i(E)}{dE}.
\end{aligned}
\end{equation}

\begin{figure}[ht!]
\centering
\includegraphics[width=0.95\columnwidth]{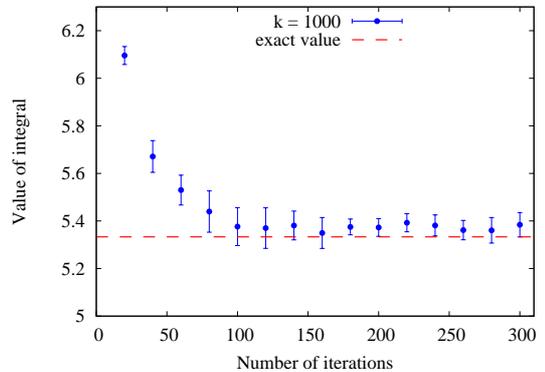}
\caption{Integral of $x^2$ over $x \in [-2,2]$, obtained using 1000
  data points in a data set and the improved scheme with a milder
  correction with the pre-factor $s = 0.25$. Error bars are obtained
  from five independent runs.}
\label{k1000_random}
\end{figure}

With these two small modifications, we revisited the integration problem
using $k = 1000$ data points in the data set (Figure
\ref{k1000_random}). The integral values in the first few dozens of
iterations deviate more from the exact value compared to the
original scheme, but it converges slowly to the exact value with a
much clearer convergence signal.  In this example, one may terminate
the simulation after e.g. the $150^{\textrm{th}}$ iteration. Another clear
improvement is that the error bar for each final answer is much reduced
compared to the original scheme, which indicates that the improved
scheme is able to give more precise results.

\section{Further improvements}

Our scheme is general in the nature that the expression of the DOS, $g(E)$,
is not restricted to the Fourier form above, as long as there is a way
to formulate the correction $c(E)$ and the update formula. Obviously,
the quality of the resulting density of states depends heavily on the
choice of an appropriate basis set. While the plane wave basis used in the
present study allowed us to implement a prototype of our algorithm
without major effort, it suffers from serious issues that will require
the choice of more suitable basis sets. In particular, local
improvements to the density of states in a limited region of the
domain should not introduce changes in regions far away. Also, the
density of states often spans a wide range of values; indeed the
density of states for the integration example possesses a singularity
at the domain boundary. Thus a more suitable, localized, basis set
will greatly improve the convergence and accuracy of our method. 
 
Although we present our method as a serial algorithm so far, we stress
its parallelization is conceptually straight-forward. Since the
generation of data points (which is the energy evaluations for a
physical system) can be done independently by distributing the work
over different processors, simple ``poor-man's'' parallelization
strategy would already guarantee significant speedup in both strong
and weak scaling.

\section{Conclusions}

In this paper, we have presented a new Monte Carlo algorithm for calculating 
probability densities of systems with a continuous energy domain. The idea is 
inspired by combining ideas from the works of Wang and Landau 
\cite{wang_efficient_2001, wang_determining_2001}, Berg and Neuhaus 
\cite{berg_multicanonical_1991,berg_multicanonical_1992}, as well as Berg and 
Harris \cite{berg_data_2008}. Nevertheless, our algorithm does not make use
of an explicit histogram as in traditional Wang-Landau or multicanonical
sampling. It is thus possible to avoid discrete binning of the collected data. 
This histogram-free approach allows us to obtain the estimated probability 
density, or the density of states, as a closed-form expression.

We have demonstrated the application of our algorithm to a stringent
test case, numerical integration. Even with the sub-optimal Fourier
sine and cosine basis sets, our current algorithm is already capable
of giving reasonable results. An important point to note is that our
algorithm requires much \textit{fewer} number of Monte Carlo steps to
finish a simulation. It is enabled by the novel way we proposed to
correct the estimated density of states, and thus the sampling
weights, where the random walk is directed consciously to achieve
uniform sampling. This is essential for decreasing the
time-to-solution ratio of a simulation, especially for complex systems
where the computation time is dominated by energy evaluations.

The numerical integration test case provides useful insights into
improving the algorithm. Possible improvements include the use of a
basis set with local support and parallelization over energy
calculations.  The choice of basis set will have significant influence on the convergence behavior and the correctness of the simulation, but it often depends on the system of interest. In general, choosing a basis set with compact support would help by enabling local updates to the density of states, thus avoiding artificial sampling barriers. Our ongoing work includes all the possibilities for perfecting the algorithm, as well as its application to simulations of physical systems to solve real-world scientific problems.

\section{Computer Code}

The following provides an overview of the code, a guide to running the code, and a detailed description of all program files.

\subsection{Overview of the Computer Code}
The associated computer code, written in Python 3.6, consists of a ``main'' program file - \texttt{binlessMUCA.py}, a sample input file - \texttt{input.txt}, and all necessary subroutines to reproduce the results from the original and improved versions of the sampling scheme discussed in this manuscript. In addition to Python 3.6, Numpy, Sympy, and Matplotlib must be installed in order to run the code. The code can then be run from the terminal via: \texttt{\$ python binlessMUCA.py input.txt} , where \texttt{input.txt} is an input file that is required as a command line argument. This input file must contain the following:

\begin{enumerate}[label=\roman*.,nosep]
  \item random number seed (can be specified or generated randomly)
  \item the function to integrate, $y(x)$
  \item integration limits, $a$ and $b$
  \item number of points in the data set, $k$
  \item maximum terms in the basis set expansion (maximum value of $m$ allowed)
  \item the Kolmogorov-Smirnov cutoff probability, $p$
  \item maximum number of iterations
  \item desired sampling scheme (original scheme described in Section \ref{algorithm}, or improved scheme described in Section \ref{improved_algorithm} with associated pre-factor $s$)
\end{enumerate}

One important note is that if the Kolmogorov-Smirnov probability is not satisfied within the specified maximum number of basis terms, the program will terminate and need to be restarted. In addition to the required input parameters, there are options to display certain plots during the simulation; these can be turned off or on by setting the associated variables to 0 or 1, respectively. The program outputs the random number seed and the estimate for the integral after each iteration. It terminates once the last iteration is completed.

\subsection{Description of Subroutines}
All necessary subroutines are included in \texttt{.py} files according to their functionality:
\begin{description}
  \item[\texttt{binlessMUCA.py}] is the main program file. It reads the input file and contains the main sampling loop that updates the estimate for the density of states after each iteration, and terminates after the specified maximum iteration number. 
  
  \item[\texttt{getExtrema.py}] finds the global minimum and maximum of the input function over the interval defined by the integration limits. 
  
  \item[\texttt{dataSetMethods.py}] contains subroutines to construct the basis sets for the remainder and density of states, fill the data sets by randomly sampling the input function according to the acceptance probability in Eq. (\ref{acc_prob}), and construct the empirical CDF and empirical remainder.
  
  \item[\texttt{correction.py}] contains subroutines to find the closed-form approximation for the empirical remainder at each iteration by using the Fourier transform to find expansion coefficients (and thus correction coefficients) and the Kolmogorov-Smirnov probability to terminate the expansion. This file includes both the original and improved methods of obtaining the correction. 
  
  \item[\texttt{KS.py}] contains the subroutine to compute the Kolmogorov-Smirnov probability following the computer code in \cite{berg_data_2008}, which uses the asymptotic expansion from \cite{stephens_use_1970}. See Appendix A for details regarding the KS test.
  
  \item[\texttt{integrate.py}] computes the integral based on the current estimate of the density of states. 
  
  \item[\texttt{plotMethods.py}]  contains subroutines to plot certain physical quantities during the simulation.

\end{description}
 
\section*{Appendix A: The Kolmogorov-Smirnov Test}

The Kolmogorov-Smirnov (KS) statistical test 
\cite{KolmogorovAN1933,Smirnoff1939} is used as the criterion to determine the 
number of terms in $c(E)$ and to terminate an iteration. Thus, it is essential 
to understand its behavior and to be aware of various subtleties in order to 
quantify its effects on the accuracy of the final results and to avoid errors. 

All variants of the KS test return a probability, $P$, that the maximum 
distance between two cumulative distribution functions (CDF's), $\Delta = 
\sup|F_1(x)-F_2(x)|$, exceeds a predetermined value $\varepsilon$ 
\cite{berg_2004}. Nevertheless, there are different ways to define and compute 
$P$.

The first distinction gives rise to the one- and two- sided tests. The one-
sided test will give two slightly different results depending on how the 
empirical CDF is defined - whether the data points are counted in the previous 
step or the next step. This amounts to computing how much the CDF's differ in 
one direction - either above or below. The two-sided test uses the absolute 
maximum deviation, regardless of direction. In other words, it combines the 
results from both one-sided KS tests and returns the smaller probability. In 
our work we have used the two-sided test. 

There exists an asymptotic expansion for the CDF of the two-sided KS statistic 
when the number of data points $k \rightarrow \infty$. If the two functions 
$F_1(x)$ and $F_2(x)$ are from the same distribution, the probability of the 
quantity $\sqrt{k} \Delta$ being smaller than or equal to a certain value $z$ 
is \cite{KolmogorovAN1933}:
\begin{equation}
\lim_{k\rightarrow \infty}\mathbb{P}(\sqrt{k}\Delta \leq z) = 1 - 2\sum_{j=1}^{\infty}{(-1)^{j-1}e^{-2j^2 z^2}}.
\end{equation}
Substituting $z = \sqrt{k} \varepsilon$, we obtain the asymptotic expansion for the probability of $\Delta$ being larger than $\varepsilon$:
\begin{equation}
\begin{aligned}
P &= \lim_{k\rightarrow \infty}\mathbb{P}(\Delta > \varepsilon) 
   = 1 - \lim_{k\rightarrow \infty}\mathbb{P}(\Delta \leq \varepsilon) \\
&= 2\sum_{j=1}^{\infty}{(-1)^{j-1}e^{-2j^2 \varepsilon^2 k}}.
\end{aligned}
\end{equation}

To apply the above results to our problem, we reverse the logic and reason that 
if we observe the maximum difference in the CDF's, $\Delta_\textrm{obs}$, from 
our data points, i.e., $\varepsilon = \Delta_\textrm{obs}$, $P$ can then be 
interpreted as the probability (level of confidence) of $F_1$ and $F_2$ coming 
from the same distribution \cite{owen_1962}. A modification to the formula that 
replaces $k$ with a modified form $S_k$ was later introduced for the 
computation of the KS statistic instead of the use of numerical tables 
\cite{stephens_use_1970}. Therefore \footnote{We note that in Ref. 
\cite{berg_2004}, there are errors with the formulae in Eq. (2.148) on Page 98. 
The implementation in the Fortran code, \texttt{kolm2\_as.f}, is nevertheless 
correct, which is the same as Eq. (\ref{KS_asymp}) here.},

\begin{equation}
\begin{aligned}
\label{KS_asymp}
P = 2\sum_{j=1}^{\infty}{(-1)^{j-1}e^{-2j^2\Delta_\textrm{obs}^2S_k}},
\end{aligned}
\end{equation}
where $S_k = \left( \sqrt{k} + 0.12 + \frac{0.11}{\sqrt{k}} \right)^2$. 

The second distinction also impacts the way the probability measure is 
computed. The one-sample test computes the probability that a \textit{single} 
data set of size $N_{D1}$ was drawn from a specified, continuously defined 
probability distribution. The two-sample test computes the probability that 
\textit{two} data sets of size $N_{D1}$ and $N_{D2}$, respectively, were drawn 
from the same underlying, unspecified probability distribution. The asymptotic 
expansion for the two-sided test in Eq. (\ref{KS_asymp}) depends on the number 
of data points, $k$, in the data set(s). In the case of the one-sample test, $k 
= N_{D1}$. However, the two-sample test uses an \textit{effective} number of 
points given by:

\begin{equation}
\begin{aligned}
\label{KSDistance}
k = \frac{N_{D1} \cdot N_{D2}}{N_{D1} + N_{D2}}.
\end{aligned}
\end{equation}

\begin{figure}[h!]
\centering
\includegraphics[width=0.95\columnwidth]{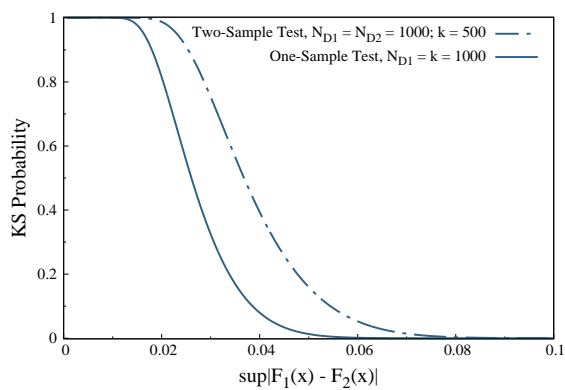}
\caption{Comparison of the probability for the one- and two-sample KS tests as a function of the maximum distance between CDF's for data set(s) of size $N_{D1} = N_{D2} = 1000$. The \textit{effective} number of data points $k$ for the two sample test is given by Eq. (\ref{KSDistance}).}
\label{KSTest1000}
\end{figure}

\begin{figure}[h!]
\centering
\includegraphics[width=0.95\columnwidth]{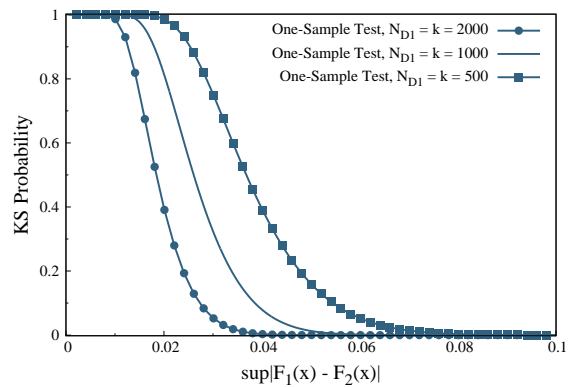}
\caption {Comparison of the one-sample KS test probability as a function of the maximum distance between CDF's for data sets of different sizes.}
\label{KSTest_N_comparison}
\end{figure}

The implications of using an effective number of points becomes apparent through analysis of the computed probability as a function of $\sup|F_1(x)-F_2(x)|$.  Figure \ref{KSTest1000} shows how the KS probability changes for the one- vs. two- sample tests for 1000 data points, where the two-sample test uses an effective number of data points, $k = 500$, given by Eq. (\ref{KSDistance}) with $N_{D1} = N_{D2} = 1000$. This figure indicates that the KS two-sample test, where each of the two data sets contains the same number of points as a single data set used with a one-sample test, results in a more lenient probability measure. Figure \ref{KSTest_N_comparison} shows how the one-sample KS probability changes as a function of the size of the data set. From this figure, it is apparent that decreasing the number of data points also decreases the stringency of the probability measure. In both figures, the solid line without symbols represents the same ``baseline'' for comparison to the other curves: a one-sample KS test with data set of size $k = 1000$. Interesting to note is that the two-sample test for two data sets of size $N_{D1} = N_{D2} = 1000$ produces exactly the same KS probability curve as does an $k = 500$ one-sample test: the dotted curve in Figure \ref{KSTest1000} and the line with squares in Figure \ref{KSTest_N_comparison} line up exactly. Thus, using the two sample test has the same effect as decreasing the size of the data set; i.e., both make the probability measure more lenient.

The above analysis shows the importance of understanding which variant of the KS test a given scientific computing package implements. The probability that the KS test returns is heavily reliant on $k$, and thus the results are heavily dependent on the particular implementation of the KS test. These subtleties must be understood in order to use the KS test correctly. The results shown in Section \ref{integration} were obtained by using the KS statistical test provided in the Mathematica software package \cite{mathematica}. However, the way to compute the probability measure is not specified in the documentation. In our Python code, we use the one-sample test. The default value of the Kolmogorov-Smirnov cutoff probability is set to be $p$ = 0.17 in order to reproduce comparable probability measure as Mathematica (which uses the two-sample test and $p$ = 0.5).

\section*{Acknowledgements}
The authors would like to thank T. W\"{u}st and D. M. Nicholson for
constructive discussions. This research was sponsored by the
Laboratory Directed Research and Development Program of Oak Ridge
National Laboratory, managed by UT-Battelle, LLC, for the
U. S. Department of Energy. This project was supported in part by an appointment to the Science Education Programs at Oak Ridge National Laboratory, administered by ORAU through the U.S. Department of Energy Oak Ridge Institute for Science and Education. This research used resources of the Oak Ridge Leadership Computing Facility, which is supported by the Office of Science of the U.S. Department of Energy under contract no. DE-AC05-00OR22725.


\bibliographystyle{elsarticle-num}
\bibliography{CPC_references}

\end{document}